\documentclass[prd,notitlepage,longbibliography,nofootinbib,superscriptaddress,onecolumn,preprintnumbers]{revtex4-2}
\usepackage[utf8]{inputenc}

\usepackage{bm}
\usepackage{comment} 
\usepackage[colorlinks=true,urlcolor=blue,anchorcolor=black,citecolor=blue,linkcolor=red,filecolor=black,menucolor=black,pagecolor=black,linktocpage=true,pdfproducer=medialab,pdfa=true]{hyperref}
\usepackage{graphicx}
\usepackage{amsmath,latexsym,amssymb,mathrsfs,ascmac,mathtools}
\usepackage{multirow}
\usepackage{braket}
\usepackage{placeins}
\usepackage{subcaption}
\begin{document}

\title{Probing Effective Field Theory Corrections with Quasinormal Modes and Gravitational Lensing in Reissner-Nordström Black Holes}

\author{Takamasa Kanai}
\email{kanai@kochi-ct.ac.jp}

\affiliation{Department of Social Design Engineering,
National Institute of Technology (KOSEN), Kochi College,
200-1 Monobe Otsu, Nankoku, Kochi, 783-8508, Japan}

\begin{abstract}
Effective field theory (EFT) provides a systematic framework for parametrizing possible higher-energy corrections to general relativity through higher-curvature interactions. In this work, we investigate gravitational lensing in both weak- and strong-field regimes for EFT-corrected Reissner-Nordström black hole spacetimes, focusing on both weakly charged and near-extremal configurations. Using the strong deflection limit formalism, we derive the corresponding corrections to the deflection angle, photon sphere radius, critical impact parameter, and strong lensing coefficients induced by higher-derivative curvature-electromagnetic interactions.

Our analysis is restricted to purely geometrical corrections associated with modifications of the background spacetime geometry, without including polarization-dependent corrections to the photon propagation law. We show that strong gravitational lensing observables in charged black hole backgrounds can provide complementary probes of effective interactions between gravity and electromagnetic fields. These results suggest that future high-precision observations of strong lensing phenomena may place constraints on higher-curvature EFT couplings beyond general relativity.
\end{abstract}
\maketitle

\section{Introduction}

Gravitational phenomena around compact objects provide an important arena for exploring possible deviations from general relativity in the strong-field regime. In particular, the behavior of null trajectories near black holes is governed by unstable circular photon orbits, which can be formulated geometrically in terms of photon surfaces \cite{Claudel:2000yi}. In static and spherically symmetric spacetimes, these structures reduce to the well-known photon sphere \cite{Virbhadra:1999nm,Claudel:2000yi}. Since photon spheres determine the propagation of light in the vicinity of black holes, they play a crucial role in various observational phenomena, including black hole shadows \cite{Falcke:1999pj,EventHorizonTelescope:2019dse,EventHorizonTelescope:2022wkp,Vagnozzi:2022moj}, quasinormal modes (QNMs) \cite{Chandrasekhar:1975zza,Iyer:1986np,Kokkotas:1999bd,Buonanno:2006ui}, and gravitational lensing \cite{Virbhadra:1999nm,Bozza:2001xd,Bozza:2002zj,Gibbons:2008rj}.

Gravitational lensing in strong field regime is particularly sensitive to the near-horizon geometry of black holes. Light rays passing close to the photon sphere experience large deflection angles, which diverge logarithmically in the strong deflection limit. This universal structure was systematically formulated by Bozza \cite{Bozza:2002zj}, where the deflection angle is separated into divergent and regular contributions. The corresponding coefficients are determined locally by the geometry near the photon sphere, making strong lensing observables useful probes of modified gravitational dynamics.

From the viewpoint of effective field theory (EFT), higher-curvature corrections provide a systematic parametrization of possible ultraviolet modifications to general relativity \cite{Weinberg:1978kz,Donoghue:1993eb,Donoghue:1994dn,Burgess:2003jk}. Such corrections generally modify the spacetime geometry surrounding compact objects and can therefore induce observable effects in strong gravitational phenomena. In charged black hole spacetimes, higher-derivative interactions involving both curvature tensors and electromagnetic fields can further alter the structure of null trajectories and consequently modify gravitational lensing observables \cite{Jing:2015kny,Chen:2015cpa}.

In previous studies \cite{Kanai:2026xpw}, strong gravitational lensing in EFT-corrected Schwarzschild spacetimes has been investigated extensively. In the present work, we extend these analyses to charged black hole backgrounds, focusing on the Reissner-Nordström spacetime with higher-curvature corrections. In particular, we analyze both weakly charged and near-extremal black holes and study how the EFT corrections affect the photon surface, QNMs and strong lensing observables. Related studies on EFT corrections to black hole perturbations and QNMs in Reissner-Nordström backgrounds have been investigated in the literature (see, e.g., Refs.~\cite{Nomura:2020tpc,Nomura:2021efi}), where the impact of higher-curvature operators on the spectrum of perturbations has been analyzed.

It should be emphasized that the present analysis is restricted to purely geometrical corrections arising from modifications of the background spacetime metric. In general, EFT interactions involving curvature couplings to electromagnetic fields can also modify the effective propagation law of photons themselves, leading to polarization-dependent effects such as gravitational birefringence. Although such effects may provide additional corrections to photon trajectories and lensing observables, they are not included in the present work. Accordingly, our analysis should be regarded as capturing only the geometrical contribution to the EFT corrections.

Another important aspect of photon spheres is their close relation to the eikonal limit of QNMs. Since both strong lensing observables and QNM spectra are controlled by unstable null orbits, they provide complementary probes of the same underlying spacetime structure. This correspondence suggests that observational signatures associated with charged black holes may contain useful information about higher-derivative interactions in gravity coupled to electromagnetic fields.

Motivated by these considerations, in this paper we investigate gravitational lensing in EFT-corrected Reissner-Nordström spacetimes. Using the strong deflection limit formalism, we compute the corresponding lensing coefficients and analyze their dependence on higher-curvature couplings. Our goal is to clarify how strong-field lensing observables encode information about effective interactions between gravity and electromagnetic fields, and to explore the extent to which future observations may constrain such EFT corrections.

In this work, we systematically investigate strong gravitational lensing in Reissner–Nordström black holes within an effective field theory framework including higher-curvature interactions. Our analysis simultaneously covers both extremal and weakly charged regimes, allowing for a unified and controlled comparison between these qualitatively distinct configurations.

The novelty of this work lies in deriving analytic expressions for the strong deflection coefficients and photon sphere quantities including higher-curvature corrections, and in clarifying how these corrections manifest differently in the near-extremal and perturbative charge regimes. This combined treatment provides a coherent framework to connect effective field theory parameters with observable strong lensing signatures.

This paper is organized as follows. In Sec.~\ref{sec:photon_surface}, we briefly review photon surfaces and their properties in static and spherically symmetric spacetimes. In Sec.~\ref{sec.EFT corrections}, we introduce the EFT framework including higher-curvature and curvature--electromagnetic interactions. In Sec.~\ref{sec.photon surface EFT}, we construct perturbative black hole solutions around the Reissner-Nordström background and discuss the resulting corrections to the photon surface. In Sec.~\ref{sec:qnm}, we analyze quasinormal modes in the eikonal approximation, while Secs.~\ref{sec:weak lens} and \ref{sec:strong lens} are devoted respectively to gravitational lensing in weak- and strong-field regimes. In Sec.~\ref{sec.propagation}, we discuss higher-curvature corrections to the photon propagation law and their implications for polarization-dependent modifications of the photon surface geometry. Finally, Sec.~\ref{sec:conclusion} contains our conclusions and future prospects.

In this paper, we set the Newton constant and the speed of light equal to unity.

\section{Photon Surface}
\label{sec:photon_surface}

In this section, we briefly review the notion of photon surfaces and their role in strong gravitational phenomena. Photon surfaces characterize the behavior of null geodesics near compact objects and play a central role in black hole shadows, quasinormal modes, and gravitational lensing \cite{Claudel:2000yi}. The photon surface is closely related to unstable circular null geodesics.
In static and spherically symmetric spacetimes, null rays propagating near the photon sphere experience large deflection angles and may orbit the black hole multiple times before escaping to infinity.
This logarithmic divergence of the deflection angle forms the basis of the strong deflection limit formalism developed by Bozza~\cite{Bozza:2002zj}.

Moreover, the properties of unstable null geodesics are also connected to the eikonal limit of quasinormal modes, where the real part of the frequency is determined by the angular velocity of the null orbit and the imaginary part is governed by its instability timescale.

A photon surface $\mathcal{S}$ is defined as a timelike hypersurface such that every null geodesic initially tangent to $\mathcal{S}$ remains tangent to it throughout its evolution \cite{Claudel:2000yi}. Geometrically, this condition is equivalent to requiring that the hypersurface be umbilical,
\begin{align}
K_{ab}\propto h_{ab},
\end{align}
where $h_{ab}$ is the induced metric on $\mathcal{S}$ and $K_{ab}$ denotes its extrinsic curvature.

For a static and spherically symmetric spacetime of the form
\begin{align}
ds^2=-A(r)dt^2+B(r)dr^2+C(r)d\Omega^2,
\end{align}
the photon surface condition reduces to
\begin{align}
\frac{d}{dr}\left(\frac{A(r)}{C(r)}\right)=0.
\end{align}
The corresponding radius determines the location of unstable circular photon orbits and therefore controls various strong-field observables. Note that, in static and spherically symmetric spacetimes, the photon surface coincides with the photon sphere, as unstable circular null geodesics are confined to a constant-radius hypersurface.

In the Schwarzschild spacetime, the photon sphere is located at $r=3M$. In the Reissner--Nordström (RN) spacetime, the metric is given by
\begin{align}
\label{metric_RN}
ds^2=-F(r)dt^2+G(r)dr^2+r^2d\Omega^2,
\qquad
F(r)=\frac{1}{G(r)}
=1-\frac{2M}{r}+\frac{Q^2}{4\pi r^2},
\end{align}
where the corresponding background electromagnetic potential is
\begin{align}
\label{potential RN}
\Phi(r)=\frac{Q}{4\pi r}.
\end{align}
The outer photon sphere radius is then determined by
\begin{align}
r_{\rm ph}
=\frac{3\pi M+\sqrt{-2\pi Q^2+9\pi^2M^2}}{2\pi}.
\end{align}

For subextremal black holes satisfying $\frac{Q^2}{4\pi}<M^2$, the outer photon sphere exists uniquely outside. The Reissner--Nordström spacetime admits two circular null orbits. Among them, the outer photon sphere is unstable and relevant for gravitational lensing observables, whereas the inner orbit lies inside the horizon for subextremal configurations and does not affect external observations.

Since the coefficients appearing in the strong deflection limit are determined locally by the geometry near the photon sphere, corrections to the spacetime geometry induced by EFT interactions directly modify gravitational lensing observables. The photon surface is closely related to unstable circular null geodesics and therefore plays a fundamental role in strong-field gravitational phenomena. In particular, light rays propagating near the photon sphere can undergo large deflection and orbit the black hole multiple times before escaping to infinity, leading to the logarithmic divergence structure characteristic of strong gravitational lensing. Furthermore, in the eikonal limit, quasinormal mode frequencies are also governed by the properties of unstable null orbits, with the real part determined by the angular velocity of the orbit and the imaginary part related to its instability timescale. Consequently, the photon surface provides a common geometrical framework connecting strong gravitational lensing, black hole shadows, and quasinormal modes.

\section{Effective Field Theory of Higher-Derivative Corrections in Einstein-Maxwell Theory}
\label{sec.EFT corrections}

In this section, we review the effective field theory (EFT) description of higher-derivative corrections to four-dimensional Einstein-Maxwell theory. Such corrections arise as low-energy effects of ultraviolet (UV) physics, including quantum gravity, string theory, or integrating out heavy fields. The EFT framework provides a systematic derivative expansion controlled by a cutoff scale $\Lambda$.

\subsection{Einstein-Maxwell Theory}

The leading-order action is given by
\begin{align}
S_0 = \frac{1}{\kappa^2} \int d^4 x \sqrt{-g} \left( R - \frac{\kappa^2}{4}F_{\mu\nu}F^{\mu\nu} \right),
\end{align}
where $\kappa^2=16\pi G$, $R$ is the Ricci scalar and $F_{\mu\nu} = \partial_\mu A_\nu - \partial_\nu A_\mu$ is the Maxwell field strength.

\subsection{Higher-Derivative Expansion}

At energies well below the cutoff scale $\Lambda$, the action is supplemented by higher-derivative operators consistent with diffeomorphism and gauge invariance. Up to four derivatives, the most general action takes the form
\begin{align}
S = S_0 + S_{\text{HD}},
\end{align}
with
\begin{align}
S_{\text{HD}} = \frac{1}{\kappa^2}\int d^4 x \sqrt{-g} \Big[
& \alpha_1 R^2
+ \alpha_2 R_{\mu\nu}R^{\mu\nu}
+ \alpha_3 R_{\mu\nu\rho\sigma}R^{\mu\nu\rho\sigma} \notag \\
& + \beta_1 R F_{\mu\nu}F^{\mu\nu}
+ \beta_2 R_{\mu\nu}F^{\mu\rho}F^{\nu}{}_{\rho}+\beta_3R_{\mu\nu\rho\sigma}F^{\mu\nu}F^{\rho\sigma} \notag \\
& + \gamma_1 (F_{\mu\nu}F^{\mu\nu})^2
+ \gamma_2 F_{\mu\nu}F^{\nu\rho}F_{\rho\sigma}F^{\sigma\mu}
\Big],
\end{align}
where the Wilson coefficients $\alpha_i$, $\beta_i$, and $\gamma_i$ are suppressed by appropriate powers of $\Lambda$.

\subsection{Reduction to Independent Operators}

Not all higher-derivative operators are physically independent. By using integration by parts, Bianchi identities, and field redefinitions, the operator basis can be reduced to a minimal set of independent interactions. In particular, in four spacetime dimensions the Gauss-Bonnet combination is topological and does not contribute to the classical equations of motion.

In the following subsection, we briefly summarize how these redundancies reduce the number of independent four-derivative operators in the Einstein-Maxwell EFT.

\subsection{Why Only a Reduced Set of Independent Operators?}

Although many four-derivative terms can be written down, several of them are not physically independent. The reduction to a minimal operator basis relies on three key observations.

\paragraph*{(i) Topological redundancy: Gauss--Bonnet term}
In four spacetime dimensions, the Gauss--Bonnet combination
\begin{align}
\mathcal{G}
=
R^2
-4R_{\mu\nu}R^{\mu\nu}
+
R_{\mu\nu\rho\sigma}R^{\mu\nu\rho\sigma},
\end{align}
is a total derivative. As a result, it does not contribute to the classical equations of motion and can be removed from the action without affecting any local dynamics. This eliminates one linear combination of curvature-squared terms.

\paragraph*{(ii) Field redefinitions and equations of motion}
Operators proportional to the leading-order equations of motion are redundant in an EFT, since they can be removed by field redefinitions of the metric and gauge field. For example,
\begin{align}
g_{\mu\nu}
&\to
g_{\mu\nu}
+
a\,R_{\mu\nu}
+
b\,R g_{\mu\nu},
\\
A_\mu
&\to
A_\mu
+
c\,\nabla^\nu F_{\nu\mu},
\end{align}
generate shifts in the higher-derivative action that allow one to eliminate terms such as
$R^2$,
$R_{\mu\nu}R^{\mu\nu}$,
$RF_{\mu\nu}F^{\mu\nu}$,
as well as mixed operators involving the Ricci tensor, in favor of other operators. Since physical observables are invariant under such redefinitions, these operators do not correspond to independent couplings.

Furthermore, terms proportional to the leading-order equations of motion do not contribute at linear order in perturbations. Expanding the action around a background solution, the leading-order equations vanish on the background, and any operator proportional to them contributes only at quadratic or higher order in perturbations. Therefore, such terms can consistently be neglected when computing linear observables.

\paragraph*{(iii) Integration by parts and identities}
Using integration by parts and Bianchi identities, some operators can be related to others or reduced to equivalent forms. This further decreases the number of independent structures that must be considered.

\paragraph*{(iv)Result and effective action}
Taking into account the above redundancies, one finds that the four-derivative Einstein--Maxwell EFT in four dimensions can be parametrized by three independent operators.

A convenient basis is then given by
\begin{align}
R_{\mu\nu\rho\sigma}F^{\mu\nu}F^{\rho\sigma},
\qquad
(F_{\mu\nu}F^{\mu\nu})^2,
\qquad
F_{\mu\nu}F^{\nu\rho}F_{\rho\sigma}F^{\sigma\mu}.
\end{align}

Therefore, the effective action up to four derivatives can be written as
\begin{align}
\label{EFT action}
S
=
\frac{1}{\kappa^2}
\int d^4x \sqrt{-g}
\Big[
R
-\frac{\kappa^2}{4}F_{\mu\nu}F^{\mu\nu}
+\alpha(F_{\mu\nu}F^{\mu\nu})^2
+\beta F_{\mu\nu}F^{\nu\rho}F_{\rho\sigma}F^{\sigma\mu}
+\gamma R_{\mu\nu\rho\sigma}F^{\mu\nu}F^{\rho\sigma}
\Big],
\end{align}
where $\alpha$, $\beta$, and $\gamma$ are the EFT coupling constants of the Einstein--Maxwell theory.

The analysis presented in this work is valid within the regime where the EFT expansion remains applicable, namely when the characteristic curvature scale of the background spacetime is sufficiently smaller than the EFT cutoff scale. Under this condition, higher-curvature terms can be consistently treated as perturbative corrections. In particular, even in the vicinity of the photon sphere, which governs strong gravitational lensing, the corrections derived in this study remain reliable as long as curvature invariants do not exceed the cutoff scale.

Accordingly, in the present work we adopt the effective action~\eqref{EFT action} as the starting point of our analysis and first investigate the corrections to the photon surface geometry induced by higher-derivative interactions. Based on these results, we then analyze how such corrections modify the QNM spectrum and gravitational lensing observables in Reissner-Nordström black hole spacetimes.

\section{Photon surface with higher curvature corrections in the Reissner-Nordström black hole spacetimes}
\label{sec.photon surface EFT}

In this section, we investigate how higher-curvature and curvature--electromagnetic interactions modify the Reissner--Nordström spacetime by solving the field equations perturbatively to first order. In particular, we compute the corresponding corrections to the photon surface and related strong-field observables.

The gravitational theory under consideration is diffeomorphism invariant, and therefore the coordinate system itself has no direct physical meaning. To provide an invariant geometrical interpretation of the radial coordinate, we fix the gauge by identifying the radial coordinate with the areal radius, so that surfaces of constant $r$ possess area $4\pi r^2$. In this gauge, the radial coordinate acquires a direct geometrical meaning, allowing us to characterize unambiguously the displacement of the photon surface induced by EFT corrections.

As discussed in the previous section, the form of the effective action is constrained by the classification of independent higher-curvature invariants in four-dimensional spacetimes. In the presence of electromagnetic fields, additional higher-derivative interactions coupling curvature tensors and the Maxwell field strength become possible. In the present work, we consider the leading corrections arising from cubic and quartic curvature invariants together with curvature--electromagnetic couplings. As discussed in the previous section, we consider the effective field theory action (\ref{EFT action}) for the Einstein-Maxwell system including higher-derivative interactions between the gravitational and electromagnetic fields. The action is given by
\begin{align}
S = \frac{1}{\kappa^2}\int d^4 x \sqrt{-g} \Big[ R 
- \frac{\kappa^2}{4} F_{\mu\nu}F^{\mu\nu}
+ \alpha \, (F_{\mu\nu}F^{\mu\nu})^2
+ \beta \, F_{\mu\nu}F^{\nu\rho}F_{\rho\sigma}F^{\sigma\mu}
+ \gamma \, R_{\mu\nu\rho\sigma}F^{\mu\nu}F^{\rho\sigma}
\Big],
\end{align}
where $\alpha$, $\beta$, and $\gamma$ are the EFT coupling constants parametrizing higher-derivative corrections to the Einstein--Maxwell theory.

We now derive the equations of motion associated with the effective action introduced above. 
Varying the action with respect to the metric yields the gravitational field equations,
\begin{align}
R_{\mu\nu}-\frac{1}{2}g_{\mu\nu}R
=
\frac{\kappa^2}{2} T_{\mu\nu}^{({\rm EM})}
+\alpha T_{1,\mu\nu}
+\beta T_{2,\mu\nu}
+\gamma T_{3,\mu\nu},
\end{align}
where the standard Maxwell energy-momentum tensor is given by
\begin{align}
T_{\mu\nu}^{({\rm EM})}
=
F_{\mu\alpha}F_{\nu}^{\ \alpha}
-\frac{1}{4}g_{\mu\nu}F_{\alpha\beta}F^{\alpha\beta},
\end{align}
and the higher-derivative contributions are
\begin{align}
T_{1,\mu\nu}
&=
4F_{\alpha\beta}F^{\alpha\beta}
\left(
F_{\mu\rho}F_{\nu}^{\ \rho}
-\frac{1}{8}g_{\mu\nu}F_{\rho\sigma}F^{\rho\sigma}
\right),\\
T_{2,\mu\nu}
&=
4F_{\alpha\mu}F_{\nu}^{\ \beta}
F_{\beta}^{\ \rho}F_{\rho}^{\ \alpha}
-\frac{1}{2}g_{\mu\nu}
F_{\alpha}^{\ \beta}F_{\beta}^{\ \rho}
F_{\rho}^{\ \sigma}F_{\sigma}^{\ \alpha},\\
T_{3,\mu\nu}
&=
3R_{\alpha\beta\rho\sigma}
\delta^\alpha_{\ (\mu}F^\beta_{\ \nu)}
F^{\rho\sigma}
+\frac{1}{2}g_{\mu\nu}
R_{\alpha\beta\rho\sigma}
F^{\alpha\beta}F^{\rho\sigma}
-2\nabla_\alpha\nabla_\beta
\left(
F^\alpha_{\ (\mu}F^\beta_{\ \nu)}
\right).
\end{align}

On the other hand, varying the action with respect to the gauge field gives the modified Maxwell equations,
\begin{align}
\nabla_\nu F^\nu_{\ \mu}
=
-\frac{8}{\kappa^2}
\Bigl[
2\alpha F_\mu^{\ \nu}
\nabla_\nu
(F_{\alpha\beta}F^{\alpha\beta})
+
2\beta F^{\nu\alpha}
\nabla_\nu
(F_{\alpha}^{\ \rho}F_{\rho\mu})
+
\gamma
\nabla_\nu
(R^{\ \nu}_{\mu\ \alpha\beta}F^{\alpha\beta})
\Bigr].
\end{align}

We investigate static perturbative corrections to the Reissner--Nordström spacetime induced by higher-derivative interactions in the effective field theory described in the previous section. In four-dimensional Einstein--Maxwell theory, the unique static and asymptotically flat charged black hole solution is given by the Reissner--Nordström geometry \cite{Misner:1973prb}, which we therefore adopt as the background spacetime.

The background metric and electromagnetic potential are given by Eqs.~(\ref{metric_RN}) and (\ref{potential RN}), respectively. We now consider static perturbations around this background and introduce a small perturbative parameter $\epsilon$. In the present analysis, higher-derivative curvature terms and curvature-electromagnetic interactions are assumed to generate corrections at order $\mathcal{O}(\epsilon)$, which are treated perturbatively. The metric and electromagnetic potential are then parametrized as
\begin{align}
\label{metric}
ds^2
&= -(F(r)+\delta F(r))dt^2
+(G(r)+\delta G(r))dr^2
+r^2\left(d\theta^2+\sin^2\theta,d\phi^2\right),\notag\\
&= -f(r)dt^2+g(r)dr^2
+r^2\left(d\theta^2+\sin^2\theta,d\phi^2\right),\\
A_{\mu}dx^\mu
&=(\Phi(r)+\delta\Phi(r))dt.
\end{align}

At linear order, the perturbation equations can generally be decomposed into independent angular modes. In the present setup, however, the background Reissner--Nordström spacetime is spherically symmetric, and the higher-derivative correction terms constructed from the background geometry and electromagnetic field inherit the same symmetry. Consequently, these higher-curvature contributions appear as spherically symmetric source terms in the perturbation equations, and the resulting first-order corrected solution also remains static and spherically symmetric. Therefore, no nonspherical perturbative modes are generated within the present framework.

This result is consistent with the uniqueness property of static, asymptotically flat charged black holes in Einstein-Maxwell theory in Ref. \cite{Yazadjiev:2015jza,Cederbaum:2015fra}, according to which the Reissner-Nordström spacetime is completely characterized by the mass and electric charge. Therefore, no nontrivial static higher-multipole perturbations arise within the present perturbative framework.

In asymptotically flat spacetimes, the physically meaningful conserved quantities are given by the ADM mass and the asymptotic electric charge, which are determined respectively from the coefficients of the $1/r$ term in the metric and the $1/r^2$ term in the electromagnetic field at spatial infinity. It is therefore convenient to absorb the corresponding perturbative corrections into redefined physical parameters,
\begin{align}
M_{\rm phys}
&=
M+\epsilon\,\delta M,\\
Q_{\rm phys}
&=
Q+\epsilon\,\delta Q.
\end{align}

Expressing the solution in terms of the physical mass and charge allows us to isolate genuine higher-order effects arising from the EFT interactions, excluding spurious contributions that merely correspond to renormalizations of the background parameters. In particular, this procedure makes it possible to identify the nontrivial corrections associated with higher-derivative curvature and curvature--electromagnetic couplings.

The resulting first-order corrections are then obtained as
\begin{align}
\delta F(r)&=-\frac{Q^4\left(2\alpha+\beta+4\pi\gamma\right)}{1280\pi^4r^6}+\frac{\gamma Q^2M}{16\pi^2r^5}-\frac{\gamma Q^2}{16\pi^2r^4},\\
\delta G(r)&=\frac{Q^2(2\alpha Q^2+\beta Q^2+16\pi\gamma(4Q^2+5\pi r(-7M+4r))}{1280\pi^4r^6\left(1-\frac{2M}{r}
+\frac{Q^2}{4\pi r^2}\right)^2},\\
\delta\Phi(r)&=-\frac{Q^3\left(2\alpha+\beta+9\pi\gamma\right)}{640\pi^2r^5}+\frac{\gamma QM}{16\pi^2r^4}.
\end{align}

The first-order perturbative solution in the present EFT setup was previously derived in Ref.~\cite{Izumi:2024rge}. We have confirmed that the solution obtained here agrees with the result presented there.

In this spacetime, the location of the photon surface is given by
\begin{align}
r_{\mathrm{ph}} =& \frac{3\pi M+\sqrt{-2\pi Q^2+9\pi^2M^2}}{2\pi}\notag\\
&-\dfrac{Q^2(3M\sqrt{\pi}+\sqrt{-2Q^2+9\pi^2M^2})^6(-2\alpha Q^2-\beta Q^2+\pi\gamma(-55M(3M+\sqrt{-8Q^2+9M^2})\pi+4Q^2(-1+30\pi)))}{20(3M+\sqrt{-8Q^2+9M^2})^9\pi^6(-2Q^2+9\pi M^2+3M\sqrt{\pi}\sqrt{-2Q^2+9\pi M^2})}.
\end{align}

In this paper, we will compute observable quantities using the perturbative solution and the location of the photon surface obtained above.

\subsection{Impact Parameter and Photon Surface}

In this subsection, we briefly review the definition of the impact parameter and its relation to the photon surface in the Reissner--Nordström spacetime including EFT corrections.

We consider a static, spherically symmetric spacetime described by the
metric given in Eq.~(\ref{metric}). For null geodesics
($ds^2=0$), restricting the motion to the equatorial plane
$\theta=\pi/2$, we obtain
\begin{align}
0
=
-f(r)\dot{t}^{\,2}
+g(r)\dot{r}^{\,2}
+r^2\dot{\phi}^{\,2}.
\end{align}
The conserved quantities can be derived from the geodesic
Lagrangian,
\begin{align}
\mathcal{L}
&=
\frac{1}{2}
\left[
-f(r)\dot{t}^{\,2}
+g(r)\dot{r}^{\,2}
+r^2\dot{\phi}^{\,2}
\right].
\end{align}
Since the metric is independent of $t$ and $\phi$, these coordinates
are cyclic, and the corresponding canonical momenta are conserved.
Thus, defining the conserved energy and angular momentum as
\begin{align}
E
&\equiv
-\frac{\partial\mathcal{L}}{\partial\dot{t}}
=
f(r)\dot{t},
\qquad
L
\equiv
\frac{\partial\mathcal{L}}{\partial\dot{\phi}}
=
r^2\dot{\phi},
\end{align}
we obtain
\begin{align}
E &= f(r)\dot{t},
\qquad
L = r^2\dot{\phi}.
\end{align}
Substituting these conserved quantities into the null condition gives
the radial equation
\begin{align}
\dot{r}^{\,2}
+
U_{\rm eff}(r)
=
\frac{E^2}{f(r)g(r)},
\end{align}
where the effective potential is defined as
\begin{align}
U_{\rm eff}(r)
=
\frac{L^2}{r^2g(r)}.
\end{align}

The impact parameter is defined as
\begin{align}
b=\frac{L}{E}.
\end{align}
Using this definition, the radial equation can be rewritten as
\begin{align}
\dot r^2=\frac{E^2}{f(r)g(r)}\left(1-\frac{b^2f(r)}{r^2}\right).
\end{align}

Circular null geodesics are determined by the conditions
\begin{align}
\dot r=0,
\qquad
\frac{dU_{\rm eff}}{dr}=0.
\end{align}
The first condition implies
\begin{align}
b^2
=
\frac{r^2}{f(r)}.
\end{align}
Evaluating this expression at the photon surface radius $r=r_{\rm ph}$ yields the critical impact parameter
\begin{align}
b_c^2
=
\frac{r_{\rm ph}^2}{f(r_{\rm ph})}.
\end{align}

The critical impact parameter determines the boundary between capture and scattering of null geodesics and therefore plays a central role in strong gravitational lensing observables. Since both the photon surface radius and the metric functions receive EFT corrections, the critical impact parameter also encodes information about higher-derivative curvature--electromagnetic interactions.
\begin{align}
b_c=&\frac{1}{2\sqrt{\pi}}\sqrt{\frac{A^4}{B}}
+\frac{\bigl(2\alpha Q^4+\beta Q^4-36\pi\gamma Q^4
  +240\pi^2\gamma Q^2M^2
  +80\pi^{3/2}\gamma Q^2M\sqrt{-2Q^2+9\pi M^2}\bigr)}
{160\,\pi^{3/2}
\Bigl(Q^2-6\pi M^2-2M\sqrt{\pi}\sqrt{-2Q^2+9\pi M^2}\Bigr)^2
\sqrt{\dfrac{C}{B}}},
\end{align}
where
\begin{align}
A=&3M\sqrt{\pi}+\sqrt{-2Q^2+9\pi M^2},\\
B=&-Q^2+6\pi M^2+2M\sqrt{\pi}\sqrt{-2Q^2+9\pi M^2},\\
C=&Q^4-36\pi Q^2M^2+162\pi^2M^4
  -6\sqrt{\pi}Q^2M\sqrt{-2Q^2+9\pi M^2}
  +54\pi^{3/2}M^3\sqrt{-2Q^2+9\pi M^2}.
\end{align}

This critical impact parameter characterizes the boundary between capture and scattering of null geodesics, and thus plays a central role in determining the strong gravitational lensing observables.

\section{Eikonal Quasinormal Modes in Effective Field Theory}
\label{sec:qnm}

Observational quantities associated with black holes may provide useful probes of effective field theory corrections to gravity. In this section, we investigate quasinormal mode (QNM) frequencies in EFT-corrected Reissner--Nordström black hole spacetimes within the eikonal approximation.

We consider a static, spherically symmetric spacetime described by the metric given in Eq.~(\ref{metric}). To study the propagation of perturbations on this background, we introduce a test scalar field $\Phi$. Although the physically relevant perturbations for gravitational waves correspond to tensor modes, the scalar-field analysis captures the same universal leading behavior in the eikonal regime and therefore provides a convenient framework for the present analysis.

The scalar field obeys the covariant Klein-Gordon equation
\begin{align}
\square\Phi
&=
\frac{1}{\sqrt{-g}}
\partial_\mu
\left(
\sqrt{-g}\,
g^{\mu\nu}
\partial_\nu\Phi
\right)\notag\\
&=
-\frac{1}{f(r)}
\partial_t^2\Phi
+
\frac{1}{\sqrt{f(r)g(r)}\,r^2}
\partial_r
\left(
\sqrt{\frac{f(r)}{g(r)}}\,r^2
\partial_r\Phi
\right)
+
\frac{1}{r^2}
\Delta_{S^2}\Phi
=0,
\label{wave equation}
\end{align}
where $g_{\mu\nu}$ and $g$ denote the background metric and its determinant, respectively. For the static and spherically symmetric metric given in Eq.~(\ref{metric_RN}), the inverse metric and the determinant are given by
\begin{align}
g^{\mu\nu}
&=
\mathrm{diag}
\left(
-\frac{1}{f(r)},
g(r),
\frac{1}{r^2},
\frac{1}{r^2\sin^2\theta}
\right),
\\
\sqrt{-g}
&=
\frac{r^2\sin\theta}
{\sqrt{f(r)g(r)}}.
\end{align}
Substituting these expressions into the covariant wave operator yields the explicit form in Eq.~(\ref{wave equation}). Here, $\Delta_{S^2}$ denotes the Laplacian operator on the unit two-sphere,
\begin{equation}
\Delta_{S^2}
=
\frac{1}{\sin\theta}
\partial_\theta
\left(
\sin\theta\,\partial_\theta
\right)
+
\frac{1}{\sin^2\theta}
\partial_\phi^2.
\end{equation}

Separating the variables using the standard scalar spherical
harmonics, we write
\begin{equation}
\Phi(t,r,\theta,\phi)
=
e^{-i\omega t}
Y_{\ell m}(\theta,\phi)
\frac{\psi(r)}{r},
\end{equation}
where $Y_{\ell m}(\theta,\phi)$ denotes the scalar spherical
harmonics, with $\ell=0,1,2,\ldots$ and
$m=-\ell,-\ell+1,\ldots,\ell$ being the angular quantum numbers.
Using the eigenvalue relation
\begin{equation}
\Delta_{S^2}Y_{\ell m}
=
-\ell(\ell+1)Y_{\ell m},
\end{equation}
the angular dependence can be separated, yielding a radial equation
for $\psi(r)$.

Introducing the tortoise coordinate
\begin{equation}
\frac{dr_*}{dr}
=
\sqrt{\frac{g(r)}{f(r)}},
\end{equation}
the radial equation can be rewritten in Schr\"odinger-like form,
\begin{equation}
\frac{d^2\psi}{dr_*^2}
+
\left(
\omega^2-V_{\rm eff}(r)
\right)\psi
=
0,
\end{equation}
with the effective potential
\begin{equation}
V_{\rm eff}(r)
=
f(r)
\left[
\frac{\ell(\ell+1)}{r^2}
+
\frac{1}{r\sqrt{f(r)g(r)}}
\frac{d}{dr}
\left(
\sqrt{\frac{f(r)}{g(r)}}
\right)
\right].
\end{equation}

\begin{figure}[t]
    \centering

    \begin{subfigure}{0.32\textwidth}
        \centering
        \includegraphics[width=\textwidth]{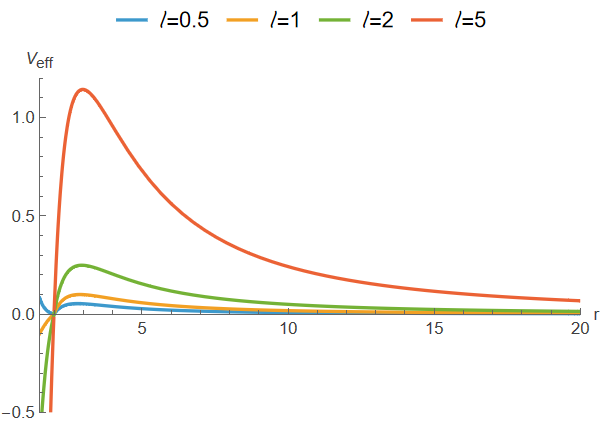}
        \caption{$Q=0.5$}
    \end{subfigure}
    \hfill
    \begin{subfigure}{0.32\textwidth}
        \centering
        \includegraphics[width=\textwidth]{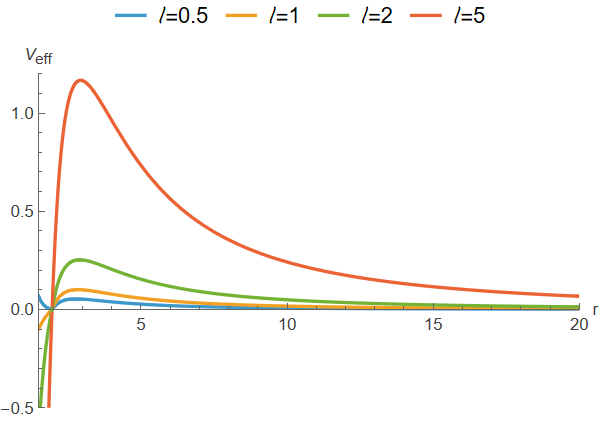}
        \caption{$Q=1$}
    \end{subfigure}
    \hfill
    \begin{subfigure}{0.32\textwidth}
        \centering
        \includegraphics[width=\textwidth]{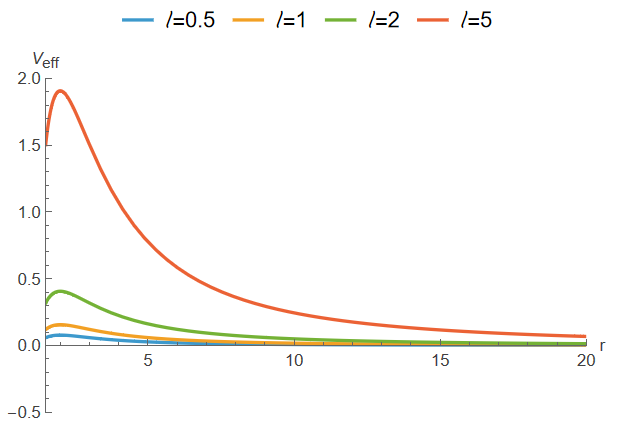}
        \caption{$Q=\sqrt{4\pi}$}
    \end{subfigure}

    \caption{Effective potential for the Reissner--Nordström black hole for different values of the charge $Q$, with $M=1$.}
    \label{fig:RN_effective_potential}
\end{figure}

\begin{figure}[t]
    \centering

    \begin{subfigure}{0.32\textwidth}
        \centering
        \includegraphics[width=\textwidth]{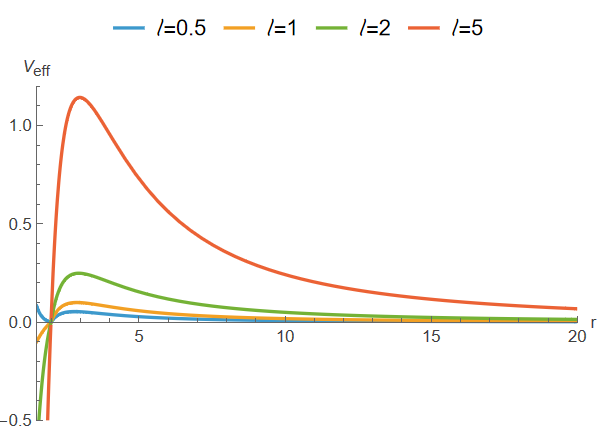}
        \caption{$Q=0.5$}
        \label{fig:RN_effective_potential_Q05}
    \end{subfigure}
    \hfill
    \begin{subfigure}{0.32\textwidth}
        \centering
        \includegraphics[width=\textwidth]{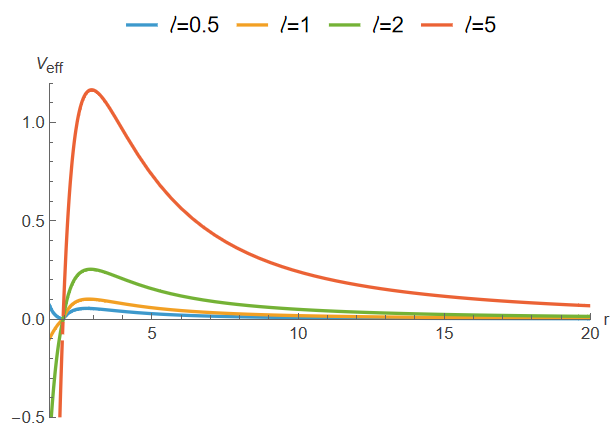}
        \caption{$Q=1$}
        \label{fig:RN_effective_potential_Q1}
    \end{subfigure}
    \hfill
    \begin{subfigure}{0.32\textwidth}
        \centering
        \includegraphics[width=\textwidth]{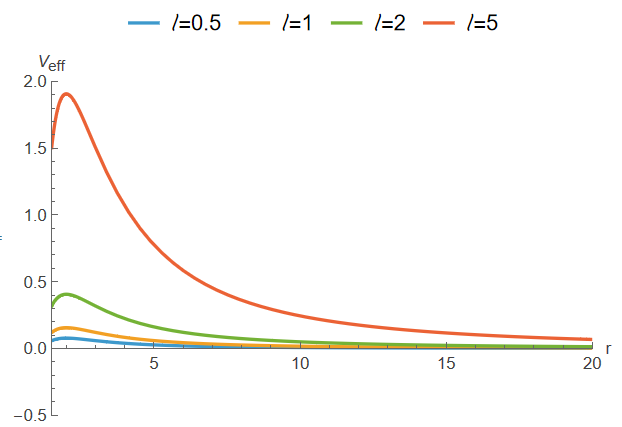}
        \caption{$Q=\sqrt{4\pi}$}
        \label{fig:RN_effective_potential_Qsqrt4pi}
    \end{subfigure}

    \caption{Effective potential for the Reissner-Nordström black hole with higher curvature correections for different values of $Q$, with $M=1$ and higher-curvature coupling constants of order $\mathcal{O}(10^{-3})$.}
    \label{fig:RN_effective_potential}
\end{figure}
Figure \ref{fig:RN_effective_potential} shows the radial dependence of the effective potential for the Reissner–Nordström black hole for $Q=0.5$, $Q=1$, and $Q=\sqrt{4\pi}$, respectively, with $M=1$. The higher-curvature coupling constants are chosen to be of order $\mathcal{O}(10^{-3})$. In all cases, the effective potential exhibits a potential barrier outside the event horizon, with its maximum located near the photon sphere. Since the higher-curvature corrections are treated perturbatively, their effects remain small for the chosen values of the coupling constants, and the corrected effective potential closely follows the standard Reissner–Nordström result. Nevertheless, the EFT corrections induce small shifts in the height and location of the potential maximum. At large radial distances, these corrections become increasingly suppressed, and the effective potential approaches the standard Reissner–Nordström behavior \cite{Media:2025udn}.

We now evaluate the QNM frequencies in the eikonal approximation for the EFT-corrected Reissner--Nordström spacetime constructed in the previous section. In the geometric optics regime, corresponding to large angular momentum,
\begin{align}
\ell\gg1,
\end{align}
wave propagation is governed by unstable circular null geodesics.

The maximum of the effective potential is determined by
\begin{equation}
\frac{d}{dr}
\left(
\frac{f(r)}{r^2}
\right)
=
0,
\end{equation}
which is precisely the condition defining the photon surface in static and spherically symmetric spacetimes.

In the eikonal limit, $\ell\gg1$, the QNM
frequencies are closely related to the properties of the unstable
circular null orbit. At leading order in the WKB approximation, the
QNM frequencies satisfy
\begin{equation}
\omega^2
\simeq
V_0
-
i\left(n+\frac12\right)
\sqrt{-2V_0''},
\end{equation}
where $V_0$ denotes the value of the effective potential at its
maximum, and $V_0''$ denotes its second derivative with respect to
the tortoise coordinate $r_*$ evaluated at the same point. Assuming
that the imaginary part of the QNM frequency is much smaller than
its real part, namely
$\left|\operatorname{Im}(\omega)\right|
\ll\operatorname{Re}(\omega)$, we obtain
\begin{equation}
\omega_{\ell n}
\simeq
\sqrt{V_0}
-
i\left(n+\frac12\right)
\sqrt{-\frac{V_0''}{2V_0}}.
\end{equation}

In the eikonal limit, the maximum of the effective potential coincides with the unstable circular null orbit. We therefore identify the location of the potential maximum with the photon-sphere radius $r_{\rm ph}$ and define
\begin{equation}
V_0 \equiv V(r_{\rm ph}),
\qquad
V_0'' \equiv
\left.
\frac{d^2V}{dr_*^2}
\right|_{r=r_{\rm ph}},
\end{equation}
where $r_*$ denotes the tortoise coordinate. Since the photon sphere
corresponds to an unstable circular null orbit, the effective
potential has a local maximum at $r=r_{\rm ph}$, implying
$V_0''<0$.

In the large-$\ell$ limit, the effective potential is dominated by
the leading centrifugal contribution proportional to
$\ell(\ell+1)$. Consequently, the leading eikonal behavior is
insensitive to the spin of the perturbing field, while
spin-dependent contributions enter only at subleading order. At the
photon sphere, the leading contribution to the effective potential
can be expressed as
\begin{equation}
V_0\simeq \ell^2\Omega_c^2,
\end{equation}
where $\Omega_c$ denotes the angular frequency of the unstable
circular null orbit. The corresponding Lyapunov exponent, which
characterizes the instability of the circular null orbit, is given
by
\begin{equation}
\lambda
\equiv
\sqrt{-\frac{V_0''}{2V_0}}.
\end{equation}
This result is consistent with the well-established correspondence between the eikonal QNMs and the properties of the unstable circular
null orbit reported in Refs.~\cite{Choudhury:2026phy,Devi:2026tsr}. Therefore, the QNM frequencies in the eikonal limit take the
standard form
\begin{equation}
\omega_{\ell n}
\simeq
\ell\Omega_c
-
i\left(n+\frac12\right)\lambda,
\end{equation}
where $n=0,1,2,\ldots$ is the overtone number.

For the EFT-corrected background considered here, evaluating the
effective potential and its second derivative at the
EFT-corrected photon-sphere radius $r_{\rm ph}$ gives
\begin{align}
V_0
={}&
\frac{4\pi\ell(\ell+1)\,B}{A^4}
+\frac{\ell(\ell+1)}{5\,A^8D}
\Bigl[
4\alpha Q^6
+2\beta Q^6
-72\pi\gamma Q^6
-18\pi\alpha Q^4M^2
-9\pi\beta Q^4M^2
+1284\pi^2\gamma Q^4M^2
\notag\\
&\quad
-4320\pi^3\gamma Q^2M^4
-6\sqrt{\pi}\alpha Q^4M\,E
-3\sqrt{\pi}\beta Q^4M\,E
+268\pi^{3/2}\gamma Q^4ME
-1440\pi^{5/2}\gamma Q^2M^3E
\Bigr],
\\
V_0''
={}&
-\frac{64\pi^2\ell(\ell+1)\,B^2D}{A^{10}}
-\frac{256\pi\ell(\ell+1)}
{5A^{20}BD}
\Bigl[
2\alpha Q^4F
+\beta Q^4
+8\pi\gamma Q^2G
\Bigr],
\end{align}
where
\begin{align}
D
={}&
-2Q^2
+9\pi M^2
+3\sqrt{\pi}M
\sqrt{-2Q^2+9\pi M^2},
\\
E
={}&
\sqrt{-2Q^2+9\pi M^2},
\\
F
={}&
4Q^{14}
-105815808\,\pi^{13/2}M^{13}
\bigl(3\sqrt{\pi}M+E\bigr)
+5038848\pi^{11/2}Q^2M^{11}
\bigl(55\sqrt{\pi}M+16E\bigr)
\notag\\
&-1038096\pi^{9/2}Q^4M^9
\bigl(93\sqrt{\pi}M+23E\bigr)
-Q^{12}
\bigl(1285\pi M^2+71\sqrt{\pi}M E\bigr)
\notag\\
&+Q^{10}
\bigl(71868\pi^2M^4
+7676\pi^{3/2}M^3E\bigr)
-108Q^8
\bigl(14593\pi^3M^6
+2271\pi^{5/2}M^5E\bigr)
\notag\\
&+648Q^6
\bigl(26241\pi^4M^8
+5309\pi^{7/2}M^7E\bigr),
\\
G
={}&
22Q^{16}
+1813985280\pi^{15/2}M^{15}
\bigl(3\sqrt{\pi}M+E\bigr)
-7558272\pi^{13/2}Q^2M^{13}
\bigl(851\sqrt{\pi}M+257E\bigr)
\notag\\
&+139968\pi^{11/2}Q^4M^{11}
\bigl(22335\sqrt{\pi}M+5983E\bigr)
-1944\pi^{9/2}Q^6M^9
\bigl(414321\sqrt{\pi}M+95771E\bigr)
\notag\\
&-Q^{14}
\bigl(7055\pi M^2+388\sqrt{\pi}M\,E\bigr)
+12Q^{12}
\bigl(34147\pi^2M^4
+3564\pi^{3/2}M^3E\bigr)
\notag\\
&-12Q^{10}
\bigl(807276\pi^3M^6
+120857\pi^{5/2}M^5E\bigr)
+648Q^8
\bigl(181868\pi^4M^8
+34887\pi^{7/2}M^7E\bigr).
\end{align}

Therefore, the eikonal QNM spectrum is determined entirely by the properties of unstable null geodesics near the photon surface. The EFT corrections modify the photon surface radius and the associated effective potential, leading to corresponding shifts in both the real and imaginary parts of the QNM frequencies.

\section{Weak-Field Gravitational Lensing in Effective Field Theory}
\label{sec:weak lens}

To establish a meaningful comparison between theory and observation, it is crucial to analyze gravitational lensing in both the weak- and strong-field regimes. These two regimes probe distinct regions of the spacetime geometry and thus offer complementary insights into the structure of the underlying gravitational theory.

In this section, we concentrate on the weak deflection regime, where light rays propagate far from the compact object and the deflection angle admits a perturbative treatment. In contrast, in the strong deflection regime, photon trajectories approach the photon sphere, where nonlinear gravitational effects dominate, resulting in large deflection angles and the emergence of relativistic images. A detailed investigation of the strong-field regime will be presented in the subsequent section. Here, we derive the relevant observables in the weak-field limit and examine their sensitivity to higher-curvature corrections.

To formulate the analysis in a more general setting, we consider a static and spherically symmetric spacetime described by
\begin{align}
ds^2 = -A(r)\,dt^2 + B(r)\,dr^2 + C(r)\,d\Omega^2,
\end{align}
where $A(r)$, $B(r)$, and $C(r)$ are arbitrary functions of the radial coordinate $r$.

For null geodesics ($ds^2=0$), restricting the motion to the
equatorial plane $\theta=\pi/2$, we obtain the conserved quantities
in the same manner as in Sec.~\ref{sec.photon surface EFT},
\begin{align}
E &= A(r)\dot{t},
\qquad
L = C(r)\dot{\phi}.
\end{align}
Using these relations, the radial equation can be written as
\begin{align}
\dot{r}^2 = \frac{E^2}{A(r)B(r)} \left[ 1 - \frac{b^2A(r)}{C(r)} \right],
\end{align}
where $r_0$ denotes the distance of closest approach, defined by the condition $\dot{r}=0$. This condition implies
\begin{align}
\frac{b^2 A(r_0)}{C(r_0)} = 1.
\end{align}

The deflection angle is then expressed as
\begin{align}
\label{deflection angle}
\alpha = 2 \int_{r_0}^{\infty} \sqrt{\frac{B(r)}{C(r)}}
\left[
\frac{C(r)}{C(r_0)} \frac{A(r_0)}{A(r)} - 1
\right]^{-1/2}dr - \pi.
\end{align}

In the weak-field regime, we expand the metric functions as
\begin{align}
A(r) = 1 + a(r), \quad
B(r) = 1 + b(r), \quad
C(r) = r^2 \left[1 + c(r)\right],
\end{align}
where $a(r)$, $b(r)$, and $c(r)$ are small quantities and are treated perturbatively.

Furthermore, we assume that the distance of closest approach satisfies $r_0 \simeq b$, where $b$ is the impact parameter. Expanding the integrand to linear order, we obtain
\begin{align}
\alpha &\simeq 2 \int_{r_0}^{\infty}
\frac{r_0 \, dr}{r \sqrt{r^2 - r_0^2}}
\left[
1
+ \frac{1}{2}\bigl(b(r) - c(r)\bigr)
- \frac{1}{2}
\frac{r^2}{r^2 - r_0^2}
\bigl(c(r) - c(r_0) + a(r_0) - a(r)\bigr)
\right] - \pi \nonumber\\
&= 2 \int_{r_0}^{\infty}
\frac{r_0 \, dr}{r \sqrt{r^2 - r_0^2}}
\left[
\frac{1}{2}\bigl(b(r) - c(r)\bigr)
- \frac{1}{2}
\frac{r^2}{r^2 - r_0^2}
\bigl(c(r) - c(r_0) + a(r_0) - a(r)\bigr)
\right].
\end{align}

\subsubsection{Straight-line approximation}

Solving the equations of motion, the perturbation functions are found to be
\begin{align}
a(r) &=-\frac{2M}{r}+\frac{Q^2}{4\pi r^2}-\frac{Q^2(Q^2(2\alpha+\beta)+4\pi\gamma( Q^2-20Mr+20r^2))}{1280\pi^4 r^6}\nonumber\\
&\sim-\frac{2M}{r}-\frac{(2\alpha+\beta)Q^4}{1280\pi^4 r^6}-\frac{\gamma Q^2}{16\pi^2 r^4},\\
b(r) &=\frac{2M}{r}-\frac{Q^2}{4\pi r^2}+\frac{Q^2(2\alpha Q^2+\beta Q^2+16\pi\gamma(4Q^2+5\pi r(-7M+4r))}{80\pi^2r^2(Q^2+4\pi r(-2M+r))^2}\nonumber\\
&\sim\frac{2M}{r}+\frac{Q^2(2\alpha Q^2+\beta Q^2+320\pi^2\gamma r^2)}{1280\pi^4r^6},\\
c(r) &=0,
\end{align}
which will be employed in the subsequent analysis.

Substituting these expressions into the deflection angle, we obtain
\begin{align}
\alpha=2\int_{r_0}^{\infty}
\frac{r_0 \, dr}{r \sqrt{r^2 - r_0^2}}
\left[\frac{1}{2}b(r)
- \frac{1}{2}\frac{r^2}{r^2 - r_0^2}\bigl(a(r_0) - a(r)\bigr)\right].
\end{align}

We first isolate the leading contribution corresponding to general relativity, and subsequently evaluate the leading corrections induced by higher-curvature interactions through an appropriate expansion of the integrand:
\begin{align}
\alpha&\sim2\int_{r_0}^{\infty}
\frac{r_0 \, dr}{r \sqrt{r^2 - r_0^2}}
\Bigl[\frac{M}{r}+\frac{Q^2(2\alpha Q^2+\beta Q^2+320\pi^2\gamma r^2)}{2560\pi^4r^6}\nonumber\\
&\ \ \ \ \ \ \ \ \ \ +\frac{1}{2}\frac{r^2}{r^2 - r_0^2}\Bigl(\frac{2M(r-r_0)}{r_0r}+\frac{(2\alpha+\beta)Q^4(r^6-r_0^6)}{1280\pi^4r_0^{6}r^6}+\frac{\gamma Q^{2}(r^4-r_0^4)}{16\pi^4r_0^4r^4}\Bigr)\Bigr]\nonumber\\
&\sim\frac{4M}{r_0}+\frac{7(2\alpha+\beta)Q^4}{8192\pi^3r_0^6}+\frac{3(1+\pi^2)\gamma Q}{64\pi^3r_0^4},
\end{align}
where the first term reproduces the standard general relativistic result, while the remaining terms encode corrections arising from higher-curvature effects.

It is worth emphasizing that contributions from higher-curvature interactions may enter at the same order as subleading terms already present within general relativity. As a result, these effects are generically intertwined with higher-order general relativistic contributions and do not appear as clean, independent leading-order signatures. Instead, their impact should be understood as deviations from the corresponding predictions of general relativity.

In the present analysis, we have retained only the leading contributions from both the general relativistic terms and the higher-curvature corrections. More rigorously, however, such corrections should be evaluated relative to the exact general relativistic result and interpreted as perturbative modifications rather than independent contributions.

\section{Strong-Field Gravitational Lensing in Effective Field Theory}
\label{sec:strong lens}

In the previous section, we analyzed gravitational lensing in the weak-field regime, where the deflection angle can be treated perturbatively. We now turn to the strong-field regime, in which light rays propagate in the vicinity of the photon sphere and nonlinear effects become significant, leading to large deflection angles and the formation of relativistic images.

\subsection{Divergent term of the deflection angle}

In this subsection, we briefly review the strong deflection limit formalism developed by Bozza \cite{Bozza:2002zj}. We introduce a new variable
\begin{align}
z = \frac{A(r) - A_0}{1 - A_0},
\end{align}
where $A_0 = A(r_0)$. The deflection angle integral (\ref{deflection angle}) can then be written as
\begin{align}
I(r_0) = \int_{0}^{1} R(z,r_0)\, f(z,r_0)\, dz,
\end{align}
with
\begin{align}
R(z,r_0) &= \frac{2 \sqrt{AB}}{C A'} (1 - A_0)\, \sqrt{C_0}, \\
f(z,r_0) &= \frac{1}{\sqrt{A_0 - \left[(1 - A_0) z + A_0\right] \frac{C_0}{C}}}.
\end{align}
Here, quantities with subscript $0$ are evaluated at $r_0$, and a prime denotes differentiation with respect to $r$.

The function $R(z, r_0)$ is regular for all $z$ and $r_0$, whereas $f(z, r_0)$ diverges in the limit $z \to 0$. To extract the leading behavior, we expand the argument of the square root in $f(z,r_0)$ up to second order in $z$:
\begin{align}
f(z,r_0) \sim f_0(z,r_0) = \frac{1}{\sqrt{p(r_0) z + q(r_0) z^2}},
\end{align}
where
\begin{align}
p(r_0) &= \frac{1 - A_0}{C_0 A'_0} \left(C'_0 A_0 - C_0 A'_0\right), \\
q(r_0) &= \frac{(1 -A_0)^2}{2 C_0^2 A'_0{}^3}
\left[
2 C_0 C'_0 A'_0{}^2
+ \left(C_0 C''_0 - 2 C'_0{}^2\right) A_0 A'_0
- C_0 C'_0 A_0 A''_0
\right].
\end{align}

If $p(r_0) \neq 0$, the divergence behaves as $z^{-1/2}$ and remains integrable. However, when $p(r_0)=0$, the divergence becomes $z^{-1}$, leading to a logarithmic divergence of the integral.

From the structure of $p(r_0)$, one finds that it vanishes at $r_0 = r_{\mathrm{ph}}$, where the photon sphere radius satisfies
\begin{align}
\label{ph radius}
\frac{d}{dr}\left(\frac{A(r)}{C(r)}\right)\bigg|_{r=r_{\mathrm{ph}}} = 0.
\end{align}
At this radius, the impact parameter reaches its critical value $b_c$, and photons follow unstable circular orbits. For $r_0 < r_{\mathrm{ph}}$, photons are captured by the black hole.

To isolate the divergence, we decompose the integral as
\begin{align}
I(r_0) = I_D(r_0) + I_R(r_0),
\end{align}
where
\begin{align}
I_D(r_0) &= \int_{0}^{1} R(0,r_{\mathrm{ph}})\, f_0(z,r_0)\, dz, \\
I_R(r_0) &= \int_{0}^{1} g(z,r_0)\, dz,
\end{align}
with
\begin{align}
g(z,r_0) = R(z,r_0) f(z,r_0) - R(0,r_{\mathrm{ph}}) f_0(z,r_0).
\end{align}

The divergent part can be evaluated analytically:
\begin{align}
I_D(r_0) = \frac{R(0,r_{\mathrm{ph}})}{2\sqrt{q(r_0)}} 
\log \left(\frac{\sqrt{q(r_0)} + \sqrt{p(r_0) + q(r_0)}}{\sqrt{p(r_0)}}\right).
\end{align}

Expanding near $r_0 = r_{\mathrm{ph}}$, we obtain
\begin{align}
p(r_0) = \frac{2 \beta_{\mathrm{ph}} A'_{\mathrm{ph}}}{1 - A_{\mathrm{ph}}} (r_0 - r_{\mathrm{ph}}) + \mathcal{O}\left((r_0 - r_{\mathrm{ph}})^2\right),
\end{align}
and
\begin{align}
q(r_{\mathrm{ph}}) = \frac{C_{\mathrm{ph}} (1 - A_{\mathrm{ph}})^2 \left(C''_{\mathrm{ph}} A_{\mathrm{ph}} - C_{\mathrm{ph}} A''_{\mathrm{ph}}\right)}{2 A_{\mathrm{ph}}^2 C'_{\mathrm{ph}}{}^2}.
\end{align}

Substituting these expressions yields
\begin{align}
I_D(r_0)= - a \log\left( \frac{r_0}{r_{\mathrm{ph}}} - 1 \right)+ b_D+ \mathcal{O}((r_0 - r_{\mathrm{ph}})\log(r_0-r_{\mathrm{ph}})),
\end{align}
with
\begin{align}
a &= \frac{R(0,r_{\mathrm{ph}})}{\sqrt{q(r_{\mathrm{ph}})}}, \\
b_D &= \frac{R(0,r_{\mathrm{ph}})}{\sqrt{q(r_{\mathrm{ph}})}}
\log \left( \frac{2(1 - A_{\mathrm{ph}})}{A'_{\mathrm{ph}} r_{\mathrm{ph}}} \right).
\end{align}

\subsection{Regular contribution to the deflection angle}

To obtain the full coefficient $b$, we must include the regular contribution following \cite{Bozza:2002zj}. Expanding $I_R(r_0)$ around $r_{\mathrm{ph}}$ gives
\begin{align}
I_R(r_0)
=
\sum_{n=0}^{\infty}
\frac{1}{n!} (r_0 - r_{\mathrm{ph}})^n
\int_{0}^{1}
\left.
\frac{\partial^n g}{\partial r_0^n}
\right|_{r_0 = r_{\mathrm{ph}}}
dz.
\end{align}

Keeping only the leading term,
\begin{align}
I_R(r_0)
=
\int_{0}^{1} g(z,r_{\mathrm{ph}})\, dz
+ \mathcal{O}(r_0 - r_{\mathrm{ph}}),
\end{align}
we define
\begin{align}
b_R = I_R(r_{\mathrm{ph}}).
\end{align}

Including the $-\pi$ contribution, we obtain
\begin{align}
b = -\pi + b_D + b_R.
\end{align}

The impact parameter is defined as
\begin{align}
\label{impact para}
u = \sqrt{\frac{C_0}{A_0}},
\end{align}
with critical value
\begin{align}
u_{\mathrm{ph}} = \sqrt{\frac{C_{\mathrm{ph}}}{A_{\mathrm{ph}}}}.
\end{align}

Expanding near $r_{\mathrm{ph}}$, we find
\begin{align}
u - u_{\mathrm{ph}}= c (r_0 - r_{\mathrm{ph}})^2+\mathcal{O}\left( (r_0 - r_{\mathrm{ph}})^3\right),
\end{align}
where
\begin{align}
c
=
\frac{C''_{\mathrm{ph}} A_{\mathrm{ph}} - C_{\mathrm{ph}} A''_{\mathrm{ph}}}{4 \sqrt{A_{\mathrm{ph}}^3 C_{\mathrm{ph}}}}.
\end{align}

Finally, the deflection angle can be expressed as
\begin{align}
\alpha(\theta)
=
- \bar{a} \log\left( \frac{\theta D_{OL}}{u_{\mathrm{ph}}} - 1 \right)
+ \bar{b},
\end{align}
with
\begin{align}
\bar{a} &= \frac{a}{2} = \frac{R(0,r_{\mathrm{ph}})}{2\sqrt{q(r_{\mathrm{ph}})}}, \\
\bar{b} &= -\pi + b_R + \bar{a} \log \left( \frac{2q(r_{\mathrm{ph}})}{A_{\mathrm{ph}}} \right).
\end{align}

\subsection{Strong-field deflection analysis for an extremal black hole ($M = Q/\sqrt{4\pi} = 1$)}

In this subsection, we analyze gravitational lensing in the strong deflection regime in the presence of higher-curvature corrections, focusing on an extremal Reissner-Nordström black hole background. Extremal black holes, defined by the saturation of the charge-to-mass bound, provide a particularly valuable theoretical setting for probing higher-curvature corrections to general relativity. In such configurations, the near-horizon geometry exhibits enhanced symmetry~\cite{Bardeen:1999px} and becomes highly sensitive to subleading contributions arising in the effective field theory expansion~\cite{Donoghue:1994dn,Strominger:1996sh}. Consequently, extremal backgrounds tend to amplify the effects of higher-derivative interactions, making them especially suitable for isolating potential deviations from classical gravity.

In contrast, weakly charged black holes are expected to exhibit only perturbative deviations from the Schwarzschild geometry, with higher-curvature effects appearing as small corrections that may be difficult to disentangle observationally. Nevertheless, they represent a more astrophysically realistic regime and therefore provide an important complementary probe.

From the perspective of gravitational lensing, extremal and weakly charged configurations can lead to qualitatively different modifications of the photon sphere structure, the critical impact parameter, and the strong deflection coefficients. While extremal black holes enhance the sensitivity to higher-curvature couplings, weakly charged cases allow for a controlled expansion and a direct comparison with the Schwarzschild limit. Therefore, a combined analysis of both regimes provides a systematic and comprehensive framework to constrain higher-curvature interactions through future high-precision observations of strong gravitational lensing~\cite{Bozza:2002zj}.

We consider the case in which the mass and charge satisfy the extremality condition $M = Q/\sqrt{4\pi}$. For notational simplicity, we further adopt the normalization $M = Q/\sqrt{4\pi} = 1$, which can be achieved by an appropriate rescaling of the radial coordinate and does not entail any loss of generality.
\begin{align}
M = \frac{Q}{\sqrt{4\pi}} = 1.
\end{align}
This choice corresponds to a rescaling of the radial coordinate and does not lead to any loss of generality. In these units, the horizon radius is located at $r = 1$, which simplifies the analytical expressions.

We consider null geodesics in a static, spherically symmetric spacetime described by the metric functions
\begin{align}
A(r)&=\left(1-\frac{1}{r}\right)^2-\frac{2\alpha+\beta+4\pi\gamma}{80\pi^2 r^6}+\frac{\gamma}{4\pi r^5}-\frac{\gamma}{4\pi r^4},\\
B(r)&=\frac{1}{\left(1-\frac{1}{r}\right)^2}+\frac{2\alpha+\beta+64\pi\gamma}{80\pi^2r^6\left(1-\frac{1}{r}\right)^4}-\frac{7\gamma}{4\pi r^5\left(1-\frac{1}{r}\right)^4}+\frac{\gamma}{\pi r^4\left(1-\frac{1}{r}\right)^4},\\
C(r)&=r^2,
\end{align}
which incorporate higher-curvature corrections up to the order considered in the effective field theory expansion around the extremal Reissner--Nordström solution.

In this background, we evaluate the deflection angle in the strong deflection limit following the formalism reviewed in the previous subsection. The functions $R(z,r_{\mathrm{ph}})$ and $f(z,r_{\mathrm{ph}})$ appearing in the integral representation of the deflection angle take the form
\begin{align} 
R(z,r_{\mathrm{ph}})&=\frac{3}{\sqrt{1+3z}} -\frac{3}{2560\,\pi^2\,(1+3z)^{3/2}} \Bigl[\alpha\bigl(-720 - 270z^3 + 736\sqrt{1+3z} + 54z^2(-65+16\sqrt{1+3z}) \notag\\ &\quad+ 12z(-309+208\sqrt{1+3z})\bigr)+\beta\bigl(-135z^3 + 27z^2(-65+16\sqrt{1+3z}) + 8(-45+46\sqrt{1+3z})\notag\\ &\quad+ 6z(-309+208\sqrt{1+3z})\bigr)+4\gamma\pi\bigl(-415 - 540z^3 + 448\sqrt{1+3z} + 72z^2(-45+16\sqrt{1+3z})\notag\\ &\quad+ 9z(-281+192\sqrt{1+3z})\bigr) \Bigr],\\
 f(z,r_{\mathrm{ph}})&=\frac{2}{\sqrt{D}}-\frac{3}{1280\,\pi^2\,D^{3/2}}\Bigl[2\alpha\bigl(-27z^4+376(-1+\sqrt{1+3z})+18z^3(-44+7\sqrt{1+3z})\notag\\ &\quad+12z(-154+107\sqrt{1+3z}) +6z^2(-410+161\sqrt{1+3z}) \bigr)+\beta\bigl( -27z^4 +376(-1+\sqrt{1+3z}) \notag\\ &\quad
+18z^3(-44+7\sqrt{1+3z}) +12z(-154+107\sqrt{1+3z}) +6z^2(-410+161\sqrt{1+3z}) \bigr)\notag\\ &+12\gamma\pi\bigl(-9z^4+z^2(265-108\sqrt{1+3z}) +z(74-92\sqrt{1+3z})+12(-1+\sqrt{1+3z}) \notag\\ &\quad +12z^3(3+\sqrt{1+3z}) \bigr) \Bigr],
\end{align} 
where $D$ is defined by 
\begin{align} 
D:=-9z^2+4(-1+\sqrt{1+3z})+6z(-3+2\sqrt{1+3z}).
 \end{align}

Substituting these expressions into the deflection angle formula and performing the integration, we obtain the regular contribution
\begin{align} 
b_{\mathrm{R}}=&-2\sqrt{2}\log\!\left(\frac{3}{8}(2+\sqrt{2})\right) +\frac{1}{1920\pi^2} \Bigl[-2\alpha\bigl(215-328\sqrt{2}+60\pi +63\sqrt{2}\log 2 -21\sqrt{2}\log 3 -21\sqrt{2}\log(2+\sqrt{2})\bigr)\notag\\ & +\beta\bigl(-215+328\sqrt{2}-60\pi -63\sqrt{2}\log 2 +21\sqrt{2}\log 3 +21\sqrt{2}\log(2+\sqrt{2})\bigr)\notag\\ & -4\gamma\pi\bigl(275-448\sqrt{2}+60\pi +198\sqrt{2}\log 2 -11\sqrt{2}\log 729 -66\sqrt{2}\log(2+\sqrt{2})\bigr) \Bigr],
\end{align}
where the full analytic expression is given above.

We next determine the strong deflection limit coefficients and photon sphere quantities. Expanding around the photon sphere radius $r_{\mathrm{ph}}$, we obtain
\begin{align} 
\beta_{\mathrm{ph}}&=\frac{9}{8}+\frac{9(6\alpha+3\beta+22\pi\gamma)}{5120\pi^2},\\ 
\bar{a}\ &=\sqrt{2}-\frac{14\alpha+7\beta+88\pi\gamma}{640\sqrt{2}\pi^2},\\
 b_{\mathrm{D}}&=2\sqrt{2}\log[3]-\frac{2\alpha(4+7\log[3])+\beta(4+7\log[3])+8\pi\gamma(12+11\log[3])}{320\sqrt{2}\pi^2},\\ 
u_{\mathrm{ph}}&=4+\frac{2\alpha+\beta+44\pi\gamma}{640\pi^2}.
 \end{align}

Combining the divergent and regular contributions, the coefficient $\bar{b}$ entering the deflection angle is given by
\begin{align} 
\bar{b}=-\pi+b_R+2\sqrt{2}\log3+\frac{\alpha(6-28\log3)+\beta(3-14\log3)-4\pi\gamma(7+44\log3)}{640\sqrt{2}\pi^2}, 
\end{align}

Finally, the deflection angle in the strong deflection limit can be written in terms of the angular position $\theta$ as
\begin{align} 
\alpha(\theta)=&-\left(\sqrt{2}-\frac{14\alpha+7\beta+88\pi\gamma}{640\sqrt{2}\pi^2}\right)\log\left(\frac{\theta D_{OL}}{4+\frac{2\alpha+\beta+44\pi\gamma}{640\pi^2}}-1\right)+2\sqrt{2}\log(8-4\sqrt{2})\notag\\ &+\frac{1}{3840\pi^2} \Bigl[ -2\alpha\bigl(430-665\sqrt{2}+120\pi +126\sqrt{2}\log 2 -42\sqrt{2}\log(2+\sqrt{2})\bigr)\notag\\ &\quad +\beta\bigl(-430+665\sqrt{2}-120\pi -126\sqrt{2}\log 2 +42\sqrt{2}\log(2+\sqrt{2})\bigr)\notag\\ &\quad-4\gamma\pi\bigl(550-875\sqrt{2}+120\pi +396\sqrt{2}\log 2 -132\sqrt{2}\log(2+\sqrt{2})\bigr) \Bigr]-\pi.
 \end{align}

The strong deflection limit corresponds to the regime in which the closest approach $r_0$ approaches the photon sphere radius $r_{\mathrm{ph}}$. In this limit, the deflection angle exhibits a logarithmic divergence, which is a universal feature of strong gravitational lensing and is here modified by higher-curvature corrections around the extremal Reissner-Nordström background.

\subsection{Strong-field deflection analysis for a weakly charged black hole with $2M=1$}

Next, we investigate gravitational lensing in the weakly charged regime. The analysis of gravitational lensing in this weak electromagnetic field limit is expected to provide information complementary to that obtained in the strongly charged, near-extremal black hole case discussed previously, particularly from the viewpoint of effective field theories involving higher-derivative curvature-electromagnetic couplings.

In the weak electromagnetic field regime, we evaluate the deflection angle in the strong deflection limit following the standard formalism. The relevant functions $R(z,r_{\mathrm{ph}})$ and $f(z,r_{\mathrm{ph}})$ appearing in the integral expression of the deflection angle are given by
\begin{align}
R(z,r_{{\mathrm{ph}}})&=2 + \frac{Q^2(1-2z)}{3\pi}+\frac{2Q^2\gamma(-1+2z+6z^2-10z^3+4z^4)}{81\pi^2}+\mathcal{O}(Q^4)\\
f(z,r_{\mathrm{ph}})&=\frac{1}{\sqrt{z^2 - \dfrac{2z^3}{3}}}
-\frac{Q^2z^2(3-10z+6z^2)}{6\sqrt{3}\,\pi\,\bigl(-z^2(-3+2z)\bigr)^{3/2}}
+\frac{Q^2\gamma\,z^2(6-53z+60z^2-9z^3-15z^4+6z^5)}
     {162\sqrt{3}\,\pi^2\,\bigl(-z^2(-3+2z)\bigr)^{3/2}}+\mathcal{O}(Q^4).
\end{align}

Substituting these expressions into the deflection angle formula and performing the integration, we obtain the regular part of the deflection angle as
\begin{align}
b_{\mathrm{R}}=&-4\operatorname{arctanh}\!\left(\frac{1}{\sqrt{3}}\right)
+2\log6
+\frac{Q^2}{243\pi^2}
\Bigl[
  54\pi\bigl(-4+\sqrt{3}+\log 12-2\log(1+\sqrt{3})\bigr)\notag\\
&+\gamma\bigl(13-4\sqrt{3}-4\log 12+8\log(1+\sqrt{3})\bigr)
\Bigr]+\mathcal{O}(Q^4).
\end{align}

We next determine the strong deflection limit coefficients and the photon sphere quantities. The corresponding expressions are given by
\begin{align}
\beta_{\mathrm{ph}}&=1+\frac{Q^2}{9\pi}-\frac{2Q^2\gamma}{243\pi^2}+\mathcal{O}(Q^4),\\
a\ &=2+\frac{2Q^2}{9\pi}-\frac{4Q^2\gamma}{243\pi^2}+\mathcal{O}(Q^4),\\
b_{\mathrm{D}}&=\log4+\frac{Q^2(3+\log4)}{9\pi}-\frac{Q^2\gamma(15+\log16)}{243\pi^2}+\mathcal{O}(Q^4),\\
u_{\mathrm{ph}}&=\frac{3\sqrt{3}}{2}-\frac{\sqrt{3}Q^2}{4\pi}+\frac{Q^2\gamma}{18\sqrt{3}\pi^2}+\mathcal{O}(Q^4).
\end{align}

Combining the divergent and regular contributions, we obtain the coefficient $\bar{b}$ appearing in the deflection angle as
\begin{align}
\bar{b}=-\pi+b_R+\log6+\frac{Q^2(2+\log6)}{9\pi}-\frac{Q^2\gamma(7+2\log6)}{243\pi^2}+\mathcal{O}(Q^4),
\end{align}

Finally, the deflection angle in the strong deflection limit can be written in terms of the angular position $\theta$ as
\begin{align}
\alpha(\theta)=&-\left(1+\frac{Q^2}{9\pi}-\frac{2Q^2\gamma}{243\pi^2}\right)\log\left(\frac{\theta D_{OL}}{\frac{3\sqrt{3}}{2}-\frac{\sqrt{3}Q^2}{4\pi}+\frac{Q^2\gamma}{18\sqrt{3}\pi^2}}-1\right)\notag\\
&+\log\left[216\left(7-4\sqrt{3}\right)\right]-\dfrac{Q^2(6-2\sqrt{3}-\log216(7-4\sqrt{3}))}{9\pi}\notag\\
&+\frac{Q^2\gamma(6-4\sqrt{3}-\log746496+8\log(1+\sqrt{3}))}{243\pi^2}-\pi+\mathcal{O}(Q^4).
\end{align}

\begin{figure}[t]
    \centering

    \begin{subfigure}{0.48\textwidth}
        \centering
        \includegraphics[width=\textwidth]{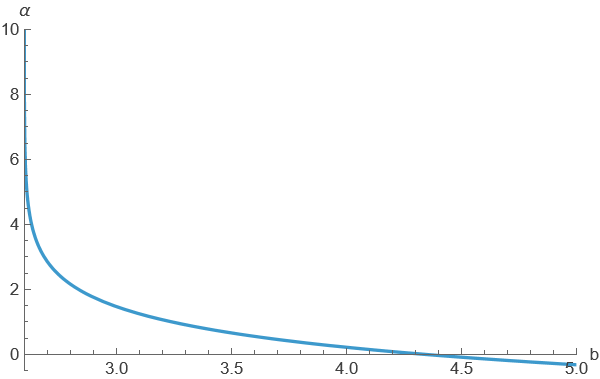}
        \caption{weak charge case, $Q=10^{-2}$}
    \end{subfigure}
    \hfill
    \begin{subfigure}{0.48\textwidth}
        \centering
        \includegraphics[width=\textwidth]{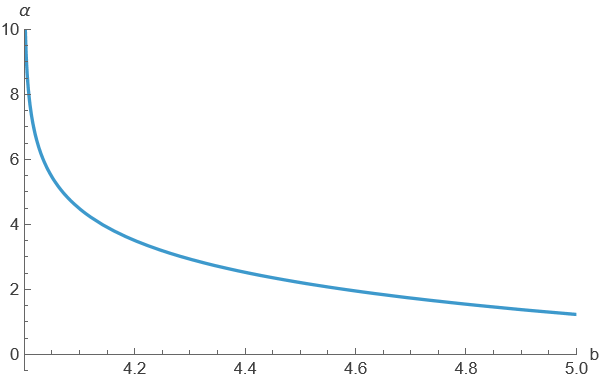}
        \caption{extremal case, $Q=\sqrt{4\pi}$}
    \end{subfigure}

    \caption{Deflection angle $\alpha(\theta)$ for the Reissner--Nordström black hole with higher-curvature corrections, with $M=1$ and higher-curvature coupling constants of order $\mathcal{O}(10^{-3})$. The left and right panels correspond to $Q=10^{-2}$ and the extremal case $Q=\sqrt{4\pi}$, respectively.}
    \label{fig:RN_deflection_angle}
\end{figure}
\begin{figure}[t]
    \centering

    \begin{subfigure}{0.48\textwidth}
        \centering
        \includegraphics[width=\textwidth]{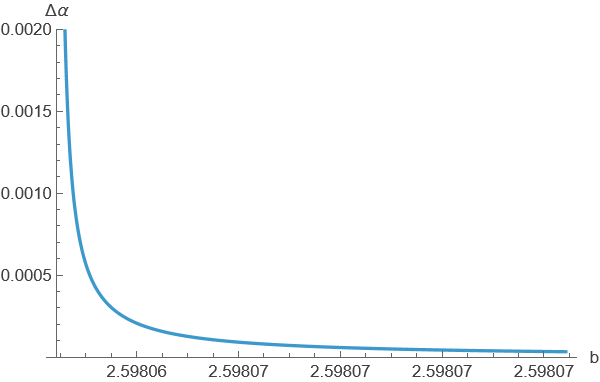}
        \caption{$Q=10^{-2}$}
    \end{subfigure}
    \hfill
    \begin{subfigure}{0.48\textwidth}
        \centering
        \includegraphics[width=\textwidth]{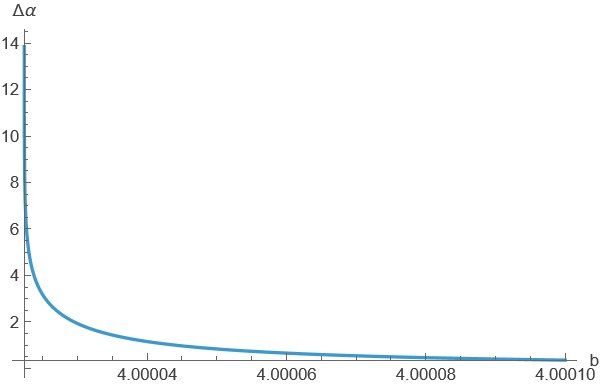}
        \caption{$Q=\sqrt{4\pi}$}
    \end{subfigure}

    \caption{Difference in the deflection angle $\Delta\alpha$ between the higher-curvature-corrected Reissner--Nordström black hole and the standard Reissner--Nordström black hole, with $M=1$ and higher-curvature coupling constants of order $\mathcal{O}(10^{-3})$. The left and right panels correspond to $Q=10^{-2}$ and the extremal case $Q=\sqrt{4\pi}$, respectively. The horizontal axis is shown with tick intervals of order $\mathcal{O}(10^{-6})$ in the impact parameter $b$.}
    \label{fig:RN_delta_deflection_angle}
\end{figure}
Since the EFT corrections are treated perturbatively, their contributions to gravitational lensing are expected to be small, making their observational signatures difficult to detect in the weak-deflection regime. However, the strong-deflection angle exhibits a logarithmic divergence as the impact parameter approaches its critical value \cite{Singh:2025tvk}. This logarithmic enhancement can amplify the otherwise small EFT contributions, potentially making them accessible to observations. This behavior can also be seen in Fig.~\ref{fig:RN_delta_deflection_angle}, where $\Delta\alpha$ becomes significantly enhanced near the critical impact parameter. Furthermore, the figure indicates that the EFT contribution becomes larger for larger values of the charge $Q$, suggesting that the effects of higher-curvature corrections become more pronounced as the black hole approaches the extremal limit.

Our analysis of gravitational lensing in charged black hole backgrounds, including both weakly charged and near-extremal Reissner--Nordström black holes, suggests that gravitational lensing observables may provide complementary information on effective field theories involving higher-derivative couplings between gravity and electromagnetic fields. In particular, the corrections obtained in the present work may offer a possible avenue for constraining such curvature-electromagnetic interactions through future high-precision observations of strong gravitational lensing phenomena.

\section{Possible Corrections to Photon Propagation and Black Hole Observables}
\label{sec.propagation}

In this work, we have focused only on the purely geometrical modifications arising from higher-curvature corrections to the background spacetime geometry. However, in effective field theories of gravity, the propagation law of photons itself is also expected to receive corrections from higher-derivative interactions. Therefore, a complete analysis of observable effects should include not only corrections to the geometry, but also modifications of the effective photon propagation. For gravitational lensing analyses incorporating such effects, see for example Ref. \cite{Chen:2015cpa,Cao:2018lrd,Zhang:2018yzr}, where these issues are discussed.

In general, higher-curvature interactions induce corrections to the effective equation governing electromagnetic waves, so that photons no longer propagate along the null geodesics of the background metric. Instead, the effective propagation law schematically takes the form
\begin{align}
(g^{\rm eff})^{\mu\nu} k_\mu k_\nu =0,
\end{align}
where $g_{\mu\nu}^{\rm eff}$ is the effective metric. In the Schwarzschild spacetime, the effective metric is explicitly given by
\begin{align}
g_{\mu\nu}^{\rm eff}=-\left(1-\frac{2M}{r}\right)dt^2+\dfrac{dr^2}{\left(1-\frac{2M}{r}\right)}+r^2\left(\frac{r^3+16\lambda M}{r^3-8\lambda M}\right)^sd\Omega^2_2.
\end{align}
Here, $\lambda$ is the coupling constant between the photon and the Weyl tensor, and the parameter $s$ is introduced to unify the expression for the two independent photon polarization modes, namely polarization along $l^\mu$ (PPL) and polarization along $m^\mu$ (PPM) (see Refs.~\cite{Jing:2015kny,Chen:2015cpa} for more details):
\begin{align}
s=
\begin{cases}
+1, & \text{for PPL},\\
-1, & \text{for PPM}.
\end{cases}
\end{align}
Here, $l^\mu$ and $m^\mu$ denote the two independent polarization modes associated with the effective metric. The polarization along $l^\mu$, referred to as the PPL mode, corresponds to the polarization lying in the plane of motion. On the other hand, the polarization along $m^\mu$, referred to as the PPM mode, corresponds to the polarization orthogonal to the plane of motion. As a consequence, the definition of the photon surface is also generalized as follows.
\begin{align}
K_{\mu\nu}\propto h_{\mu}^{\ \alpha} h_{\nu}^{\ \beta} g_{\alpha\beta}^{\rm eff}.
\end{align}

As a consequence, the location of the photon sphere (or photon surface) can receive additional corrections beyond those originating from the background geometry itself. In the geometrical optics approximation, for a perturbed static and spherically symmetric spacetime of the form
\begin{align}
ds^2=-(f(r)+\delta f(r))dt^2+\frac{dr^2}{f(r)+\delta g(r)}+r^2d\Omega^2,
\end{align}
the photon surface is determined by
\begin{align}
\dfrac{d}{dr}\left(\dfrac{f(r)+\delta f(r)}{r^2\left(\frac{r^3+16\lambda M}{r^3-8\lambda M}\right)^s}\right)=0,
\end{align}
where $\lambda$ denotes the perturbative expansion parameter associated with the coupling constant of the $R_{\mu\nu\rho\sigma}F^{\mu\nu}F^{\rho\sigma}$ term.

Therefore, once the photon propagation law is modified, this condition is expected to change, leading to corrections of the form
\begin{align}
r_{\rm ph}
\rightarrow
r_{\rm ph}+\delta r_{\rm ph}.
\end{align}

Such modifications are also expected to affect quasinormal modes (QNMs). In the eikonal limit, QNM frequencies are closely related to unstable circular photon orbits. Therefore, corrections to the effective photon propagation may induce shifts in the QNM spectrum,
\begin{align}
\omega_{\rm QNM}
\rightarrow
\omega_{\rm QNM}
+\delta\omega .
\end{align}
Similarly, gravitational lensing observables may also receive corrections through modified photon trajectories,
\begin{align}
\alpha
\rightarrow
\alpha+\delta\alpha ,
\end{align}
where $\alpha$ denotes the deflection angle.

In particular, these corrections generally depend on the polarization mode of photons. This polarization dependence significantly complicates the analysis, since different photon polarizations may effectively propagate along different characteristic cones. Consequently, phenomena such as gravitational birefringence can arise.

In Ref.~\cite{Kanai:2026xpw}, gravitational lensing was investigated in the Schwarzschild background spacetime for the effective action
\begin{align}
S=\frac{1}{16\pi G}\int d^4x\sqrt{-g}\left(R+\gamma R^{\rho\sigma}_{\ \ \mu\nu}R^{\alpha\beta}_{\ \ \rho\sigma}R^{\mu\nu}_{\ \ \alpha\beta}+\eta (R_{\mu\nu\rho\sigma}R^{\mu\nu\rho\sigma})^2+\tilde{\eta} (R_{\mu\nu\rho\sigma}\tilde{R}^{\mu\nu\rho\sigma})^2\right),
\end{align}
where the dual tensor is defined as
\begin{align}
\tilde{R}_{\mu\nu\rho\sigma}=\frac{1}{2}\epsilon_{\mu\nu}^{\ \ \alpha\beta}R_{\alpha\beta\rho\sigma}.
\end{align}
Taking into account the correction to the photon propagation law induced by the curvature-photon coupling, the photon sphere radius is modified as follows.

For the polarization mode along $m^\mu$, the photon sphere radius can be expressed in a general form as
\begin{align}
r_{\mathrm{ph}}
= 3M - \frac{4\left(704\eta + 405\gamma M^2 + 2187s \lambda M^4\right)}{6561 M^5},
\end{align}
where $s=\pm1$ depends on the choice of the polarization vector.

In particular, when we choose the polarization vector as
\begin{align}
m^a = -\frac{L}{r^2} (\partial_\theta)^a ,
\end{align}
we obtain
\begin{align}
r_{\mathrm{ph}}
= 3M - \frac{4\left(704\eta + 405\gamma M^2 -2187
\lambda M^4\right)}{6561 M^5}.
\end{align}

In the present work, we restrict our analysis to the purely geometrical contributions associated with corrections to the background metric. A systematic investigation including polarization-dependent modifications of photon propagation is left for future work. 

\section{Conclusion and Discussion}
\label{sec:conclusion}

In this paper, we have investigated weak and strong gravitational lensing in EFT-corrected Reissner-Nordström black hole spacetimes, focusing on both weakly charged and near-extremal configurations. In particular, we analyzed how higher-curvature interactions involving gravitational and electromagnetic fields modify the photon sphere geometry and the corresponding gravitational lensing observables.

Using the strong deflection limit formalism, we derived the corrections to the deflection angle, the photon sphere radius, the critical impact parameter, and the strong lensing coefficients induced by higher-derivative curvature--electromagnetic couplings. Since these quantities are determined locally by the geometry near unstable null circular orbits, they encode direct information about the underlying EFT parameters. In addition, exploiting the eikonal approximation, we estimated the quasinormal mode (QNM) frequencies associated with these spacetimes, which are likewise governed by the properties of the photon sphere.

Our analysis shows that strong gravitational lensing observables in charged black hole backgrounds can provide complementary probes of effective interactions between gravity and electromagnetic fields. In particular, the dependence of the strong deflection coefficients on the corrected photon sphere geometry suggests that future high-precision observations of black hole shadows and strong lensing phenomena may place constraints on higher-curvature EFT couplings beyond general relativity. The parallel sensitivity of eikonal QNM frequencies to the same geometric structure further strengthens this connection between strong-field observables and EFT parameters.

A key result of this work is the derivation of explicit analytic corrections to the photon sphere radius, critical impact parameter, strong deflection coefficients, and the corresponding eikonal QNM frequencies induced by higher-curvature interactions in charged black hole spacetimes.

By analyzing both extremal and weakly charged regimes within the same effective field theory framework, we have demonstrated that these two configurations exhibit qualitatively different sensitivities to higher-curvature couplings: extremal black holes amplify such effects due to their near-horizon structure, while weakly charged cases provide a controlled perturbative expansion around the Schwarzschild limit.

This unified treatment highlights how strong gravitational lensing observables, together with eikonal QNMs, can serve as complementary probes of higher-derivative interactions, and establishes a systematic bridge between effective field theory parameters and potentially observable strong-field signatures.

Although the present work has focused on purely geometrical corrections associated with modifications of the background spacetime geometry, effective field theories generally also induce corrections to the photon propagation law itself through curvature-electromagnetic couplings. As discussed in the previous section, such effects may lead to polarization-dependent photon trajectories, modifications of the photon sphere condition, and gravitational birefringence. A systematic analysis including these polarization-dependent corrections remains an important problem for future work.

Several further directions also remain to be explored. In particular, extending the present analysis to rotating black hole spacetimes would significantly enhance its astrophysical relevance. It would also be important to compare the theoretical predictions obtained here with future observational data from black hole imaging and strong gravitational lensing experiments, as well as gravitational wave observations probing QNMs. Furthermore, investigating higher-order corrections in the strong deflection expansion may improve the accuracy of the EFT predictions in the strong-field regime.

Overall, our results suggest that gravitational lensing in charged black hole spacetimes, together with eikonal QNMs, provides a promising observational window into higher-curvature interactions in gravity coupled to electromagnetic fields. Combined with complementary observables such as QNMs and black hole shadows, strong-field lensing phenomena may offer a useful avenue for probing effective field theory corrections beyond general relativity.
\bibliography{references}

\end{document}